\documentclass[useAMS,usenatbib]{mn2e}

\usepackage{graphicx}                   
\input{epsf.sty}                        
\input{psfig.sty}
\begin{document}

\title[PSR B1828$-$11: a precession pulsar torqued by a quark planet?]
{PSR B1828$-$11: a precession pulsar torqued by a quark planet?}

\author[Liu, Yue, Xu]{K. Liu$^{1}$, Y. L. Yue$^{2}$, and R. X. Xu$^{2}$\\
$^{1}$Department of Geophysics, School of Earth and Space Science, Peking University, China\\
$^{2}$Department of Astronomy, School of Physics, Peking University,
China}

\maketitle

\begin{abstract}
The pulsar PSR B1828$-$11 has long-term, highly periodic and
correlated variations in both pulse shape and the rate of
slow-down. This phenomenon may provide evidence for precession of
the pulsar as suggested previously within the framework of free
precession as well as forced one. On a presumption of forced
precession, we propose a quark planet model to this precession
phenomenon instead, in which the pulsar is torqued by a quark
planet. We construct this model by constraining mass of the pulsar
($M_{\rm psr}$), mass of the planet ($M_{\rm pl}$) and orbital
radius of the planet ($r_{\rm pl}$). Five aspects are considered:
derived relation between $M_{\rm psr}$ and $r_{\rm pl}$, movement
of the pulsar around the center of mass, ratio of $M_{\rm psr}$
and $M_{\rm pl}$, gravitational wave radiation timescale of the
planetary system, and death-line criterion. We also calculate the
range of precession period derivative and gravitational wave
strength (at earth) permitted by the model. Under reasonable
parameters, the observed phenomenon can be understood by a pulsar
($10^{-4}\sim10^{-1}M_{\odot}$) with a quark planet
($10^{-8}\sim10^{-3}M_{\odot}$) orbiting it. According to the
calculations presented, the pulsar would be a quark star because
of its low mass, which might eject a lump of quark matter (to
become a planet around) during its birth.
\end{abstract}

\begin{keywords}
Pulsars: individual (PSR B1828$-$11) --- stars: planetary systems
--- gravitational waves
\end{keywords}

\section{Introduction}
The pulsar PSR B1828$-$11 shows long-term, highly periodic and
correlated variations in both the pulse shape and the slow-down
rate. Its variations are best described as harmonically related
sinusoids, with periods of approximately 1000, 500 and 250 days
\cite[]{Stairs00}. The phenomenon indicates the most compelling
evidence for precession \cite[]{Link01}.

To explain this phenomenon, some authors
\citep[]{Jones01,Rezania03} have proposed different models within
the framework of free precession. The observation could not be a
problem in the standard view of neutron stars if the star's crust
is free to precess. In \cite{Link01}, the correlated changes in
the pulse duration and spin period derivative can be explained as
a precession of the star's rigid crust coupled to the magnetic
dipole torque. \cite{Akgun06} modelled the timing behavior with
the inclusion of both geometrical and spin-down contributions to
the residuals. However, investigations concerned on internal
structure of neutron stars show that free precession may be damped
out if vortices pinning to the stellar crust and hydrodynamics
(MHD) coupling are taken into consideration. In detail, the
rotation of the superfluid, accounting for a large proportion of
the moment of inertia of the pulsar, is contained in an array of
vortices. Models, in which vortices pinned to the stellar crust
become unpinned during a glitch, might have described the
occurrence of and recovery from glitches \cite[]{Alpar84}. The
vortex pinning will damp out free precession on timescales of
several hundred precession periods \citep[]{Shaham77,Sedrakian99}
if the pinning force is as strong as suggested in the glitch
models. Additionally, the MHD coupling between the crust and the
core will also strongly affect precession of the pulsar
\cite[]{Levin04}. The decay of precession, caused by the mutual
friction between the neutron superfluid and the plasma in the
core, is expected to occur over tens to hundreds of precession
periods and may be measurable over a human lifetime.
As noted by \cite{Link03,Link06}, the picture of vortex lines
entangled in flux tubes appears to be incompatible with
observations of long-period precession, which indicates that the
standard scenario of the outer core (superfluid neutrons in
co-existence with type II, superconducting protons) should be
reconsidered.

An alternative way is to consider the pulsar as a solid quark star
\cite[]{Xu03}, where precession models will not receive puzzle that
damping out brings. But there are still some problems when here we
come to the model of free precession. For example, the ellipticity
(or dynamical flattening) of the pulsar derived from free precession
model is not fitted well with the one calculated by Maclaurin
approximation. Consider the pulsar as a rotation ellipsoid with the
principal moment of inertia $I_{x}=I_{y}<I_{z}$ and the
corresponding radii $a=b>c$. In free precession models, the stellar
dynamical flattening is
$\epsilon=(I_{z}-I_{x})/I_{x}=e^{2}/(2-e^{2})=P/P_{\rm
prece}\approx10^{-8}$, where $P$ is the spin period, $P_{\rm prece}$
is the precession period of the pulsar and $e=\sqrt{1-c^{2}/a^{2}}$
is the stellar eccentricity. It can be approximated as $e^{2}/2$
since $e\ll1$. Meanwhile, from Maclaurin approximation, the stellar
ellipticity can be determined by
$\varepsilon=[1-(c/a)^{2/3}]/(c/a)^{2/3}\approx
e^{2}/3\approx2\epsilon/3\approx3\times10^{-3}P_{\rm
10ms}^{-2}\approx2\times10^{-6}$ \citep[]{Xu06,Zhou04}. These two
values, $\epsilon$ and $\varepsilon$, which are expected to be
generally matched if free precession model is available, are quite
different. Therefore, new ideas need to be devised to explain the
phenomenon of precession instead of the free precession ones.
Actually, a forced precession model driven by an fossil disk was
presented in \cite{Qiao03}.\footnote{In this paper, the authors
obtained the stellar oblateness from Maclaurin approximation and
considered the precession as the whole star's motion, not only the
crust's. So we think that here an idea of solid quark star is better
than neutron star so as to prevent decay of precession.}

Here we alternatively present a quark planet model to explain the
phenomenon of precession. In this model, forced precession is caused
by a quark planet orbiting the pulsar. In Sect. 2, first we
establish the relation between mass of the pulsar and orbital radius
of the planet in case that the dynamical flattening is obtained from
Maclaurin approximation \cite[]{Xu06}. Then we explain why the
planet should be a quark planet, rather than a normal one like the
earth or Jupiter when planet model is referred to. Next we limit
movement of the pulsar around the center of mass by errors in the
TOAs (time-of-arrival). Death-line criterion and limitation on
gravitation wave radiation timescale are also considered so as to
constrain orbital radius, mass of the pulsar and mass of the planet.
In Sect. 3, we calculate the precession period derivative and
gravitational wave radiation strength of the pulsar for different
mass of the pulsar and orbital radius of the planet. In Sect.4, we
conclude by discussing the formation of such system and expecting
further observation to test the model.

\section{Precession torqued by a quark planet}
\begin{figure}
\includegraphics[width=3in]{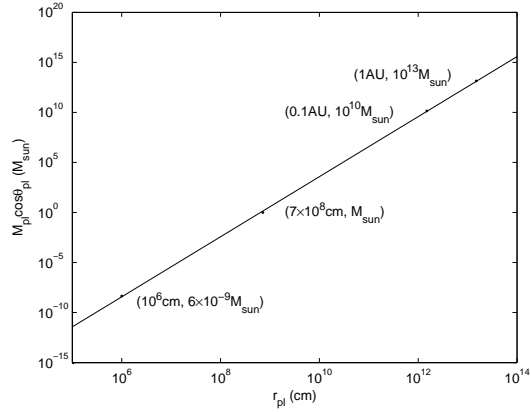}
\caption{ Relation between mass of the planet ($M_{\rm pl}$) and
orbital radius ($r_{\rm pl}$) for 500-day precession period from Eq.
(2). We can have $M_{\rm pl}\cos\theta_{\rm pl}\approx M_{\rm pl}$
if $\theta_{\rm pl}$ is not close to $90^{\circ}$. The figure
indicates that reasonable value of $M_{\rm p}$ ($M_{\rm
pl}<M_{\odot}$) may be found while $r_{\rm pl}$ is less than
$10^{9}$ cm.
\label{1}}
\end{figure}

First of all, we suppose that the pulsar PSR1828$-$11 could be
either neutron star or quark star since both are candidate models
for pulsar. In case that the precession period is much more than the
spin period and the orbital period, the angular velocity of forced
precession can be expressed as \cite[]{Menke04}
\begin{equation}
\dot{\alpha}=\frac{3GM_{\rm pl}}{2\omega r_{\rm
pl}^{3}}\epsilon\cos\theta_{\rm pl},
\label{Mr1}
\end{equation}
where $\dot{\alpha}$ is the precession angular velocity, $\omega$ is
the rotational angular velocity of the spinning pulsar, $r_{\rm pl}$
is orbital radius of the planet, $G$ is gravitational constant,
$M_{\rm pl}$ is mass of the planet,
$\epsilon\approx3\varepsilon/2\approx3\times10^{-6}$ is the stellar
dynamical flattening and $\theta_{\rm pl}$ is average inclination of
planet orbit. From Eq. (1) we can have
\begin{equation}
M_{\rm pl}=\frac{8\pi^{2}}{3GPP_{\rm prece}\epsilon\cos\theta_{\rm
pl}}r_{\rm pl}^{3},
\label{Mr2}
\end{equation}
where $P=2\pi/\omega_{\rm p}$ is spin period and $P_{\rm
prece}=2\pi/\dot{\alpha}$ is precession period of the pulsar. If we
consider a normal planet similar to the earth or Jupiter, the
typical value of $r_{\rm pl}$ should be 0.1 AU or 1 AU and the
corresponding value of $M_{\rm pl}$ is much larger than the solar
mass. In Fig. 1, the relation between $M_{\rm p}\cos\theta_{\rm pl}$
and $r_{\rm pl}$ is shown derived from the 500-day precession
period. We can see that if $r_{\rm pl}$ reaches $10^{9}~\rm cm$ and
$\theta_{\rm pl}$ is not close to $90^{\circ}$, mass of the planet
will be over the solar mass. A planet needs to be of several billion
times of $M_{\odot}$ to provide enough torque if it locates at 1 AU
away from the pulsar. We cannot believe the existence of such a
planet since it would definitely induce huge orbital timing effects
in the pulse residuals. However, the result is not surprising
because the pulsar has a much shorter forced precession period and
thus the torque that dominates the precession needs to be much
stronger. Meanwhile, precession torque is reduced when the distance
between the pulsar and the planet becomes longer. Consequently, if
there is a planet close to the pulsar, it may be able to provide
enough torque to cause the short-period precession. That is the
reason that we consider quark planet since its orbital radius could
mainly depend on the kick energy, which can vary in a large range
(Sect. 4).

Therefore, we suppose that the orbital radius of the planet is
between ($10^{6}$ cm, $10^{9}$ cm), where $10^{6}$ cm is typical
radius of a normal neutron star. Besides, for PSR B1828$-$11, the
errors in the TOAs (time-of-arrival) are limited by random noise to
about $\tau_{\rm c}\approx$ 0.2 ms \cite[]{Stairs00}. So in such a
planetary system, the pulsar is not likely to move more than about
$\tau_{\rm c}c\approx6\times10^{6}$ cm around the center of mass,
and its orbital radius should be less than $3\times10^{6}$ cm. In
addition, if the eccentricity of the orbit is not considerable, we
can have $M_{\rm pl}r_{\rm pl}\approx M_{\rm psr}r_{\rm psr}$, where
$M_{\rm psr}$ and $r_{\rm psr}$ are mass of the pulsar and its
orbital radius around the center of mass, respectively. Since
$r_{\rm psr}$ has a maximum $r_{\rm psr,\rm max}$=$3\times10^{6}$
cm, the relation can be derived as $M_{\rm pl}r_{\rm pl}< M_{\rm
psr}r_{\rm psr,\rm max}$. Finally, in such a planetary system, mass
of the pulsar should be much larger than that of the planet so we
approximately assume $M_{\rm psr}/M_{\rm pl}>k=10$.

\begin{figure}
\includegraphics[width=3in]{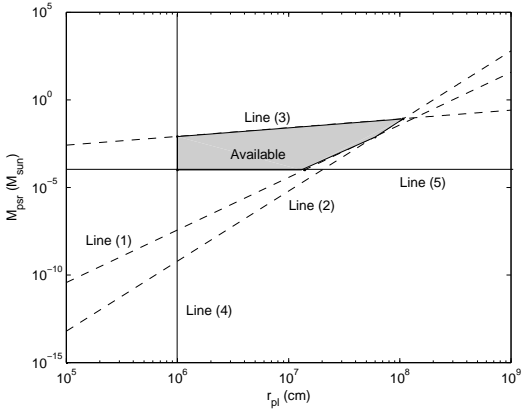}
\caption{ Constraint on $r_{\rm pl}$ (the orbital radius of the
planet) and $M_{\rm psr}$ (the mass of the pulsar) by observations
and theoretical arguments. The shadowed `Available' region
surrounded by Line (1)--(5) is the parameter space for $r_{\rm pl}$
and $M_{\rm psr}$. The five lines are defined by Eq. (20)--(24). In
this figure we use $\cos\theta_{\rm pl}=1$, since the locations of
Line (1)--(3) will not change much with variation of average
inclination of planet orbit $\theta_{\rm pl}$ from $0^{\circ}$ to
$80^{\circ}$ (see Table 1). \label{2}}
\end{figure}

Next we consider the gravitational wave radiation (GWR) of the
planetary system for further limitation. In normal double neutron
star system, the distance between the two stars is about $10^{10}$
cm. Therefore the timescale of GWR is rather long, usually $10^{4}$
years \citep[]{Hulse75,Taylor82}. However, in this quark planet
system, due to the short distance between the two objects, the power
of GWR may be significantly larger. In double-star system it is
given by ~\cite[]{Misner73}
\begin{equation}
\frac{{\rm d}E}{{\rm
d}t}=\frac{32G^{4}}{5c^{5}}\frac{\mu^{2}m^{3}}{a^{5}}=\frac{32G^{4}M_{\rm
psr}^{2}M_{\rm pl}^{2}(M_{\rm psr}+M_{\rm pl})}{5c^{5}a^{5}},
\label{EMr1}
\end{equation}
where $a\simeq r_{\rm pl}$ is the semi-major axis of the orbit,
$m=M_{\rm pl}+M_{\rm psr}$ and $\mu=M_{\rm pl}M_{\rm psr}/(M_{\rm
pl}+M_{\rm psr})$ is the reduced mass. The GWR costs the total of
potential energy and dynamic energy of the planetary system
\begin{equation}
E_{\rm tot}=-\frac{GM_{\rm pl}M_{\rm psr}}{2r_{\rm pl}}.
\label{EMr2}
\end{equation}
So timescale of GWR can be derived as
\begin{equation}
\tau=\frac{|E_{\rm tot}|}{{\rm d}E/{\rm d}t}=\frac{5c^{5}r_{\rm
pl}^{4}}{64G^{3}M_{\rm pl}M_{\rm psr}(M_{\rm psr}+M_{\rm pl})}.
\label{Mt}
\end{equation}
If the planetary system is stable, the timescale must be long
enough. Here we approximately set it as $\tau>\tau_{0}=10^{4}$
years$\approx3\times10^{13}$ s.

Additionally, we consider the death-line criterion, which requests
the potential drop at the polar cap of the pulsar be more than
$\phi_{0}\simeq10^{12}$ V \citep[]{Ruderman75,Usov95}. If we assume
that PSR1828$-$11 is an aligned pulsar, potential drop is
\begin{equation}
\phi\approx\frac{\pi R^{2}B}{cP}\sin^{2}\theta,
\label{B1}
\end{equation}
where $B$ is the polar magnetic field strength at pulsar surface,
$R\approx(3M_{\rm psr}/(4\pi\rho))^{1/3}$ is the pulsar radius,
$\rho\approx7\times10^{14}~\rm g/cm^{3}$ is the density of the
pulsar and $\theta=\arcsin\sqrt{2\pi R/(cP)}$ is the opening
half-angle of the polar cap. Here we use the density for quark stars
to obtain the lower limit of $M_{\rm psr}$. The magnetic field can
be approximated by \cite{Manchester77}
\begin{equation}
B\approx\sqrt{\frac{3Ic^{3}P\dot{P}}{8\pi^{2}R^{6}}},
\label{B2}
\end{equation}
where $I\approx(2/5)M_{\rm psr}R^{2}$ is the principal moment of
inertia. From Eq. (6) and (7), the relation between potential drop
and mass of the pulsar can be derived as below
\begin{equation}
\phi\approx(\frac{3}{5})^{1/2}(\frac{3\pi^{2}}{4})^{1/3}(\frac{\dot{P}}{cP^{3}})^{1/2}(\frac{1}{\rho})^{1/3}M_{\rm
psr}^{5/6}. \label{M}
\end{equation}
While $\phi>\phi_{0}$, we have
\begin{equation}
M_{\rm
psr}>(\frac{2000c^{3}\rho^{2}P^{9}\phi_{0}^{6}}{243\pi^{4}\dot{P}^{3}})^{1/5}\approx3\times10^{-3}M_{\odot}.
\label{M2}
\end{equation}
Actually, the assumption of alignment in PSR1828$-$11 is rather
strong. The potential drop from Eq. (6) can be larger by more than
one order of magnitude if the inclination of the magnetic axis to
the spin axis is not zero \cite[]{Yue06}. Consequently, constraint
on mass of the pulsar can be lower by about one magnitude and thus
we have $M_{\rm s}>10^{-4}M_{\odot}$.

Now there are five limitations for $M_{\rm psr}$, $r_{\rm pl}$ and
$M_{\rm pl}$:
\begin{eqnarray}
M_{\rm psr}/M_{\rm pl}>k,  \\
M_{\rm pl}r_{\rm pl}< M_{\rm psr}r_{\rm psr,\rm max}, \\
\tau=\frac{5c^{5}r_{\rm pl}^{4}}{64G^{3}M_{\rm pl}M_{\rm psr}(M_{\rm
s}+M_{\rm pl})}>\tau_{0},  \\
r_{\rm pl}\in(10^{6}~\rm cm,~10^{9}~\rm cm),  \\
M_{\rm psr}>10^{-4}M_{\odot}.
\end{eqnarray}
If we consider $M_{\rm psr}\gg M_{\rm pl}$ and substitute for
$M_{\rm pl}$ in term of $r_{\rm pl}$ according to Eq. (2), then the
limitations can be derived as
\begin{eqnarray}
~M_{\rm psr}>\frac{8k\pi^{2}}{3PP_{\rm
prece}G\epsilon\cos\theta_{\rm
pl}}r_{\rm pl}^{3},\\
~M_{\rm psr}>\frac{8\pi^{2}}{3PP_{\rm
prece}G\epsilon\cos\theta_{\rm pl}r_{\rm psr,\rm max}}r_{\rm pl}^{4},\\
~M_{\rm psr}<(\frac{15PP_{\rm
prece}c^{5}\epsilon}{512\pi^{2}G^{2}\tau_{0}\cos\theta_{\rm
pl}}r_{\rm pl})^{1/2},\\
10^{6}~\rm cm<r_{\rm pl}<10^{9}~\rm cm,\\
M_{\rm psr}>10^{-4}M_{\odot}.
\end{eqnarray}

In Fig. 2 we consider the above limitations and figure out the
available range for $M_{\rm psr}$ and $r_{\rm pl}$. Accordingly,
point ($r_{\rm pl}$, $M_{\rm psr}$) should be above Line (1) and
(2), below Line (3) and (5) and on the right of Line (4). So we have
the shadowed area as the "Available" area for point ($r_{\rm pl}$,
$M_{\rm psr}$). Line (1)-(5) are defined as below
\begin{eqnarray}
{\rm Line~(1)}:~M_{\rm psr}=\frac{8k\pi^{2}}{3PP_{\rm
prece}G\epsilon\cos\theta_{\rm pl}}r_{\rm pl}^{3},    \\
{\rm Line~(2)}:~M_{\rm psr}=\frac{8\pi^{2}}{3PP_{\rm
prece}G\epsilon\cos\theta_{\rm pl}r_{\rm psr,\rm max}}r_{\rm pl}^{4},    \\
{\rm Line~(3)}:~M_{\rm psr}=(\frac{15PP_{\rm
prece}c^{5}\epsilon}{512\pi^{2}G^{2}\tau_{0}\cos\theta_{\rm
pl}})^{1/2}r_{\rm pl}^{1/2},    \\
{\rm Line~(4)}:~r_{\rm pl}=10^{6}~\rm cm,            \\
{\rm Line~(5)}:~M_{\rm psr}=10^{-4}M_{\odot}.
\end{eqnarray}
From Fig. 2, the available value range of $r_{\rm pl}$, $M_{\rm
psr}$ and $M_{\rm pl}$ are ($10^{6}$ cm,
$5\times10^{7}\cos^{1/7}\theta_{\rm pl}$ cm), ($10^{-4}M_{\odot}$,
$6\times10^{-3}\cos^{-3/7}\theta_{\rm pl}M_{\odot}$) and
($6\times10^{-9}\cos^{-1}\theta_{\rm pl}M_{\odot}$,
$6\times10^{-4}\cos^{-4/7}\theta_{\rm pl}M_{\odot}$). Here we use
$\cos\theta_{\rm pl}=1$ because the positions of Line (1), (2) and
(3) do not vary distinctly with the changing of $\theta_{\rm pl}$
from $0^{\circ}$ to $80^{\circ}$. Different value ranges of $r_{\rm
pl}$, $M_{\rm psr}$ and $M_{\rm pl}$ with different $\theta_{\rm
pl}$ are shown in Table 1.

\begin{table*}
\centering
\begin{minipage}{140mm}
      \caption{The parametric range of $r_{\rm pl}$, $M_{\rm psr}$ and $M_{\rm pl}$ for
      different inclination of planet orbit, $\theta_{\rm pl}$. The variation of the range
      is not significant for $\theta_{\rm pl}$ from $0^{\circ}$ to $80^{\circ}$.}
      \begin{tabular}{c c c c}
            \hline\hline
            $\theta_{\rm pl}$ &~~~~~~~~~~~$r_{\rm p}$ &$M_{\rm s}$ &~~~~~~~~~~~$M_{\rm p}$\\

            \hline

            $0^{\circ}$ &($10^{6}$ cm, $5\times10^{7}$ cm)
&($10^{-4}M_{\odot}$, $6\times10^{-3}M_{\odot}$)
&($6\times10^{-9}M_{\odot}$, $6\times10^{-4}M_{\odot}$) \\
$30^{\circ}$ &($10^{6}$ cm, $5\times10^{7}$ cm)
&($10^{-4}M_{\odot}$, $6\times10^{-3}M_{\odot}$)
&($7\times10^{-9}M_{\odot}$, $7\times10^{-4}M_{\odot}$)\\
$60^{\circ}$  &($10^{6}$ cm, $5\times10^{7}$ cm)
&($10^{-4}M_{\odot}$, $8\times10^{-3}M_{\odot}$)
&($1\times10^{-8}M_{\odot}$, $9\times10^{-4}M_{\odot}$)\\
$80^{\circ}$  &($10^{6}$ cm, $4\times10^{7}$ cm)
&($10^{-4}M_{\odot}$, $1\times10^{-2}M_{\odot}$)
&($3\times10^{-8}M_{\odot}$, $2\times10^{-3}M_{\odot}$)\\

            \hline
\end{tabular}
\end{minipage}
\end{table*}

\section{To test the model by further observation}
Loss of the total energy of the system caused by gravitational wave
radiation will lead to decay of the planet orbit. Correspondingly,
precession period will be reduced with the decreasing of orbital
radius. Meanwhile, the planetary system may act as a detectable
gravitational wave source. Therefore, it is possible to test and
improve the model by GWR detection and long-period observation for
precession period derivative ($\dot{P}_{\rm prece}$).

So next we calculate $\dot{P}_{\rm prece}$ and the characteristic
amplitude of GWR source strength ($h_{\rm c}$) for different $r_{\rm
pl}$ and $M_{\rm psr}$, respectively. From Eq. (2), if orbital
radius has a slight change, the variation of $P_{\rm prece}$ can be
expressed as
\begin{equation}
\Delta P_{\rm prece}=\frac{8\pi^{2}r_{\rm pl}^{2}}{GPM_{\rm
pl}\epsilon\cos\theta_{\rm pl}}\Delta r_{\rm pl}.
\label{PM}
\end{equation}
In this case we do not consider the change of spin period (The
result will prove its reasonableness). Similarly, from Eq. (4) we
have
\begin{equation}
\Delta E_{\rm tot}=\frac{GM_{\rm pl}M_{\rm psr}}{2r_{\rm
pl}^{2}}\Delta r_{\rm pl},
\label{EM}
\end{equation}
In addition, the slight change of mechanical energy is caused by GWR
in a short period (see Eq. [3])
\begin{equation}
\Delta E_{\rm tot}=\frac{32G^{4}M_{\rm psr}^{2}M_{\rm pl}^{2}(M_{\rm
psr}+M_{\rm pl})}{5c^{5}r_{\rm pl}^{5}}\Delta t. \label{EM2}
\end{equation}
Considering $M_{\rm psr}\gg M_{\rm pl}$ and combining Eq. (25)--(27)
give the period derivative of precession as below
\begin{equation}
\dot{P}_{\rm prece}=\frac{\Delta P_{\rm prece}}{\Delta
t}=\frac{512\pi^{2}G^{2}M_{\rm
psr}^{2}}{5c^{5}P\epsilon\cos\theta_{\rm pl}r_{\rm pl}}.
\label{Pt}
\end{equation}
Meanwhile, rate of loss of angular momentum caused by GWR is
\cite{Ushomirsky00}
\begin{equation}
N_{\rm gw}=\frac{\dot{E}}{\Omega}=\frac{c^{3}\Omega d^{2}h_{\rm
a}^{2}}{G}, \label{Eh}
\end{equation}
where $\dot{E}$ is the rate of loss of the total energy,
$\Omega=\sqrt{GM_{\rm psr}/r_{\rm pl}^{3}}$ is the period of
revolution of the planet and $d$=3.58 kpc \cite[]{Taylor93} is the
distance of the pulsar. $h_{\rm a}$ is the source's `angle-averaged'
field strength (at earth) and approximately we have $h_{\rm
a}\approx h_{\rm c}$ ($h_{\rm a}\approx1.15h_{\rm c}$, see
\cite{Ushomirsky00}). Combining Eq. (2), (3) and (29) gives
\begin{equation}
h_{\rm a}=\frac{32\sqrt{2}\pi^{2}G}{3\sqrt{5}dc^{4}PP_{\rm
prece}\epsilon\cos\theta_{\rm pl}}M_{\rm psr}r_{\rm pl}^{2}.
\label{hM}
\end{equation}
Besides, the frequency of GWR is
\begin{equation}
\nu=2\frac{\Omega}{2\pi}=\sqrt{\frac{GM_{\rm psr}}{\pi^{2}r_{\rm
pl}^{3}}}. \label{fM}
\end{equation}
\begin{figure}
\includegraphics[width=3in]{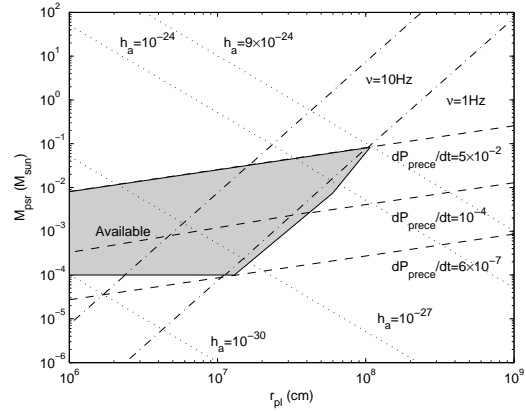}%
\caption{ A zoomed parameter space for the `Available' region. The
procession period derivative ($\dot{P}_{\rm prece}={\rm d}P_{\rm
prece}/{\rm d}t$, dash lines), the perturbed metric ($h_{\rm a}$,
dot lines), and gravitational wave frequency ($\nu$, dash-dot lines)
are drawn. Here we use $\cos\theta_{\rm pl}=1$ in the calculations.
The available area gives model-permitted parameter space for
$\dot{P}_{\rm prece}$, $h_{\rm a}$ and $\nu$.  \label{3}}
\end{figure}

In Fig. 3, relations between $r_{\rm pl}$ and $M_{\rm pl}$ from Eq.
(28) for a group of $\dot{P}_{\rm prece}$, from Eq. (31) for a group
of $h_{\rm a}$ and from Eq. (32) for a group of $\nu$ are shown
($\cos\theta_{\rm pl}\simeq1$). The relations are limited by the
available area for point ($r_{\rm pl}$, $M_{\rm pl}$) from Fig. 1.
As is shown, the maximum and minimum of $\dot{P}_{\rm prece}$ are
$4\times10^{-4}$ and $6\times10^{-7}$ while those of $h_{\rm a}$ are
$3\times10^{-25}$ and $1\times10^{-30}$. The result indicates that
the precession period changes much quickly than spin period of the
pulsar and GWR at earth is not intense enough to be detected by LIGO
at its working frequency. For example, at a frequency of 10 Hz,
value of $h_{\rm a}$ is about $10^{-27}$, which is below the current
detection limit of the LIGO at the same frequency (about
$10^{-22}$).

\section{Conclusion and Discussion}

Within the framework of forced precession, we propose a quark
planet model to explain the precession of PSR B1828$-$11. The
observed phenomenon can be understood by a pulsar (probably a
quark star) together with a quark planet which torques dominantly
the pulsar to precess. In principle, orbital radius of the quark
planet should be between $10^{6}$ cm and $10^{8}$ cm while the
range of mass of the pulsar and the planet are approximately
($10^{-4}M_{\odot}$, $10^{-1}M_{\odot}$) and ($10^{-8}M_{\odot}$,
$10^{-3}M_{\odot}$), respectively. These results might not be
strange since other candidates of low-mass quark stars were also
discussed previously \citep[]{Xu05,Yue06}. We calculate the
model-permitted precession period derivative and characteristic
amplitude of GWR for the system. The precession period changes
much quickly than spin period of the pulsar; meanwhile, GWR
strength at earth may not be large enough to be detected by
current LIGO.

If there is a quark planet providing torque for the forced
precession of pulsar PSR B1828$-$11, it should be close to the
pulsar with a distance of several times of the pulsar's radius.
The pulsar mass should also be significantly lower than $M_\odot$,
which may suggest that the pulsar would be a quark star. Such kind
of planets, orbiting closely to the center pulsars, could be
ejecta during the formation of the quark stars with strong
turbulence if the surface energy is reasonably low \cite[]{Xu06}.
Considering the orientation of the system's angular momentum, the
planet is not likely to have a inclination of orbit very close to
$90^{\circ}$ and our previous analysis with $\theta_{\rm pl}$
varying from $0^{\circ}$ to $90^{\circ}$ can work effectively.

In this paper we do not consider the possibility of more than one
planet, which may provide a way to explain the other two possible
precession periods of the pulsar. The limitation on orbital radius
and ratio of the pulsar mass to the planet one could be improved
if the formation of the system is considered. However, precession
periods of the pulsar cannot be exactly obtained now from the
seemingly periodic post-fit timing residuals. Long-period timing
observations in the future are necessary in order to obtain more
accurate precession period derivative of the system. Since 2000,
the pulsar has accomplished several precession period. Therefore,
if precession period derivative reaches its maximum in this model,
the precession period may have changed several days. In addition,
we need observation for gravitational wave to test and improve the
model. Whether it can be detected or not will both provide further
limitation on mass of the pulsar and orbital radius of the planet.

The model could also be tested by X-ray observation.
(1) If the pulsar is a solid quark star with $M_{\rm
psr}\sim10^{-3}M_{\odot}$ and then $r_{\rm psr}\sim$2~km, the rate
of rotation energy loss could be only about $10^5$ times smaller
than that in the standard model where the pulsar is a normal
neutron star.
Assuming that only $\sim 0.1\%$ of the spin-down power could turn
into the non-thermal X-ray luminosity \cite[]{Lorimer05}, the flux
at earth should be about $\rm 1\times10^{-17}erg\cdot cm^{-3}\cdot
s^{-1}$ (less than 1 photon/60 hours).
(2) The thermal X-ray emissivity from the pulsar could also be
lower. If thermal emission is from the global star, the flux
should be $\sim 5\times 10^{-13}$$\rm erg\cdot cm^{-3}\cdot
s^{-1}$ and $\sim 3\times10^{-14}$ $\rm erg\cdot cm^{-3}\cdot
s^{-1}$ for the pulsar with surface temperatures of 200 eV and 100
eV, respectively. Taking absorption into consideration, we can
expect a flux of $\sim 10^{-14}$ $\rm erg\cdot cm^{-3}\cdot
s^{-1}$ (about 70 photons/6 hours).
But if the pulsar is a neutron star with radius 10 km and surface
temperature $>60$ eV,\footnote{%
According to the standard cooling model by \cite{Page98}, the
temperature of this $\sim 10^5$ years old pulsar is $\sim 68$ eV.
} %
the flux is much higher, $> 10^{-13}$ $\rm erg\cdot cm^{-3}\cdot
s^{-1}$.
Future observations of the pulsar by Chandra or XMM-Newton could
certainly bring us more details about the real nature.

Finally, we note that the nature of pulsars (to be neutron or
quark stars) is still a matter of debate even after 40 years of
the discovery. The reason for this situation is in both theory
(the uncertainty of non-perturbative nature of strong interaction)
and observation (the difficulty to distinguish them).
It is a non-mainstream idea that pulsars are actually quark stars,
but this possibility cannot be ruled out yet according to either
first principles or observations.
``Low-mass'' is a natural and direct consequence if pulsars are
quark stars since quark stars with mass $< 1 M_\odot$ are
self-confined by color interaction rather by gravity.
An argument against the low-mass idea could be the statistical
mass-distribution of pulsars in binaries ($\sim 1.4 M_\odot$).
However this objection might not be so strong due to \citep{Xu05}:
(1) if the kick energy is approximately the same, only solar-mass
pulsars can survive in binaries as low-mass pulsars may be ejected
by the kick; (2) low-mass bare strange stars might be uncovered by
re-processing the timing data of radio pulsars if the pulsars' mass
is not conventionally supposed to be $\sim 1.4M_\odot$.
In this work, we just try to understand the peculiar precession
nature of PSR B1828-11 in the quark star scenario, since the
mainstream-scientific solution to precession might not be simple and
natural.


{\em Acknowledgments}: We would like to thank the referee for
his/her very constructive suggestion to test the model by X-ray
observations, and thank K. J. Lee and G. J. Qiao for their help
and appreciate various stimulating discussions in the pulsar group
of Peking University. This work is supported by National Nature
Sciences Foundation of China (10573002, 10778611) and by the Key
Grant Project of Chinese Ministry of Education (305001).


\end{document}